\documentclass[12pt]{article}
 \usepackage{pstricks}

\def\a{\alpha}
\def\b{\beta}

\def\e{\epsilon}
\def\f{\phi}
\def\g{\gamma}

\def\l{\lambda}
\def\m{\mu}
\def\n{\nu}

\def\p{\pi}

\def\G{\Gamma}

\def\S{\Sigma}

\def\vf{\varphi}

\def\cl{{\cal L}}
\def\cm{{\cal M}}

\def\co{{\cal O}}

\def\bo{{\raise.15ex\hbox{\large$\Box$}}}               
\def\pr{\prod}                                          
\def\face{{\raise.2ex\hbox{$\displaystyle \bigodot$}\mskip-2.2mu \llap {$\ddot
        \smile$}}}                                      
\def\dg{\dagger}                                     

\def\slash#1{\rlap{\hbox{$\mskip 1 mu /$}}#1}      
\def\wt#1{\widetilde{#1}}                    
\def\Bar#1{\overline{#1}}                       
\def\leftrightarrowfill{$\mathsurround=0pt \mathord\leftarrow \mkern-6mu
        \cleaders\hbox{$\mkern-2mu \mathord- \mkern-2mu$}\hfill
        \mkern-6mu \mathord\rightarrow$}       
\def\dvec#1{\vbox{\ialign{##\crcr
        \leftrightarrowfill\crcr\noalign{\kern-1pt\nointerlineskip}
        $\hfil\displaystyle{#1}\hfil$\crcr}}}           

\def\beq{\begin{equation}}
\def\eeq{\end{equation}}

\def\beqx{\begin{displaymath}}
\def\eeqx{\end{displaymath}}

\def\beqa{\begin{eqnarray}}
\def\eeqa{\end{eqnarray}}
\def\NO{\nonumber}

\def\pl#1#2#3{Phys.~Lett.~{\bf B {#1}} (19{#2}) #3}
\def\np#1#2#3{Nucl.~Phys.~{\bf B {#1}} (19{#2}) #3}

\def\pr#1#2#3{Phys.~Rev.~{\bf D {#1}} (19{#2}) #3}
\def\zp#1#2#3{Z.~Phys.~{\bf C {#1}} (19{#2}) #3}

\catcode`@=11
\def\@citex[#1]#2{\if@filesw\immediate\write\@auxout{\string\citation{#2}}\fi
  \def\@citea{}\@cite{\@for\@citeb:=#2\do
    {\@citea\def\@citea{,\penalty\@m}\@ifundefined
      {b@\@citeb}{{\bf ?}\@warning
       {Citation `\@citeb' on page \thepage \space undefined}}%
\hbox{\csname b@\@citeb\endcsname}}}{#1}}

\def\citer{\@ifnextchar [{\@tempswatrue\@citexr}{\@tempswafalse\@citexr[]}}
\def\@citexr[#1]#2{\if@filesw\immediate\write\@auxout{\string\citation{#2}}\fi
  \def\@citea{}\@cite{\@for\@citeb:=#2\do
    {\@citea\def\@citea{--\penalty\@m}\@ifundefined
       {b@\@citeb}{{\bf ?}\@warning
       {Citation `\@citeb' on page \thepage \space undefined}}%
\hbox{\csname b@\@citeb\endcsname}}}{#1}}
\catcode`@=12

\begin{document}
\date{\mbox{ }}

\title{ 
{\normalsize     
DESY 97-190\hfill{\tt ISSN 0418-9833}\\
April 1998\hfill{\tt hep-ph/9710460}\\
REVISED VERSION\hfill\mbox{ }}\\[25mm]
\bf CP ASYMMETRY IN\\ MAJORANA NEUTRINO DECAYS\\[8mm]}
\author{W.~Buchm\"uller and M.~Pl\"umacher\\
Deutsches Elektronen-Synchrotron DESY, 22603 Hamburg, Germany}
\maketitle

\thispagestyle{empty}

\begin{abstract}
  \noindent
  We study CP asymmetries in lepton-number violating two-body
  scattering processes and show how they are related to CP asymmetries
  in the decays of intermediate massive Majorana neutrinos. 
  Self-energy corrections, which do not contribute to CP asymmetries in 
  two-body processes, induce CP violating couplings of the
  intermediate Majorana neutrinos to lepton-Higgs states. We briefly comment 
  on the implications of these results for applications at finite temperature. 
\end{abstract}

\newpage

\noindent
{\bf\large Introduction}\\
  
  Decays of heavy Majorana neutrinos may be responsible for most of
  the cosmological baryon asymmetry \cite{fy}. As detailed studies
  have shown, the observed asymmetry $n_B/s \sim 10^{-10}$ is
  naturally obtained for theoretically well motivated patterns of
  neutrino masses and mixings, without \cite{luty,pluemi,bp} and with
  \cite{camp,pluemi2} supersymmetry.
  
  CP asymmetries in heavy particle decays are conventionally
  evaluated from the interference between tree diagrams and 
  one-loop vertex corrections \cite{kt1}. In addition interference terms
with self-energy corrections have been considered for several models
\cite{liu,flanz}, which may have large effects in some cases \cite{pil}. 
However, the correct treatment of self-energy contributions for a decaying
particle is not obvious. The naive prescription leads to a well-defined
result for the CP asymmetry, yet the individual partial decay widths are
infinite.

In this paper we shall investigate this problem in the case of
  heavy Majorana neutrinos, which are obtained as mass eigenstates
  if right-handed neutrinos are added to the standard model. Since they 
  are unstable, they cannot appear as in- or out-states of S-matrix
  elements. Rather, their properties are defined by appropriate
  S-matrix elements for stable particles \cite{veltman}. For such
  scattering processes one may define CP asymmetries for which the resonance
  contributions, at least in some approximation, can be used to define
 CP asymmetries for decays of the intermediate unstable particles. 

For applications at finite temperature the separation of two-body scattering
processes in resonance contributions and remainder is crucial \cite{wolfram}.
We shall work this out in the case of heavy Majorana 
neutrinos using a resummed propagator for the intermediate heavy neutrinos. 
It turns out that for CP asymmetries of two-body processes various
cancellations occur. The asymmetry for the full propagator vanishes 
identically\footnote{In a previous version of this paper we 
concluded incorrectly that therefore self-energy corrections can be neglected 
at finite temperature.}. Away from resonance poles, the entire CP asymmetry
vanishes to leading order \cite{rcv}. This further emphasizes the importance
to analyse CP asymmetries of two-body processes in the resonance region. 
In the following we shall study this in detail and comment on possible
implications at finite temperature.\\   

\noindent
{\bf\large Self-energy and vertex corrections}\\

  We consider the standard model with three additional right-handed
  neutrinos. The corresponding Lagrangian for Yukawa couplings and
  masses of charged leptons and neutrinos reads
  \beq
    \cl_Y = \overline{l_L}\,\f\,\l^*_l\,e_R
            +\overline{l_L}\,\wt{\f}\,\l_{\n}^*\,\n_R
            -{1\over2}\,\overline{\n^C_R}\,M\,\n_R
            +\mbox{ h.c.}\;,
  \eeq
  where $l_L=(\n_L,e_L)$ is the left-handed lepton doublet and
  $\f=(\vf^+,\vf^0)$ is the standard model Higgs doublet. $\l_{l}$,
  $\l_{\n}$ and $M$ are $3\times3$ complex matrices in the case of
  three generations.  One can always choose a basis for the fields
  $\n_R$ such that the mass matrix $M$ is diagonal and real with
  eigenvalues $M_i$. The corresponding physical mass eigenstates are
  then the three Majorana neutrinos $N_i={\n_{R}}_i + {\n_R^C}_i$. At
  tree level the propagator matrix of these Majorana neutrinos reads
  \beq
    i\,S_0(q)\,C^{-1}={i\over\slash{q}-M+i\e}\,C^{-1}\;,\label{S0}
  \eeq  
  where $C$ is the charge conjugation matrix.  This propagator has
  poles at $q^2=M_i^2$ corresponding to stable particles, whereas the 
 physical Majorana neutrinos are unstable. This is taken into account by
  summing self-energy diagrams in the usual way, which leads to the
  resummed propagator
  \beq
    i\,S(q)\,C^{-1}={i\over\slash{q}-M-\S(q)}C^{-1}\;.\label{S}
  \eeq

\begin{figure}[t]
    \input{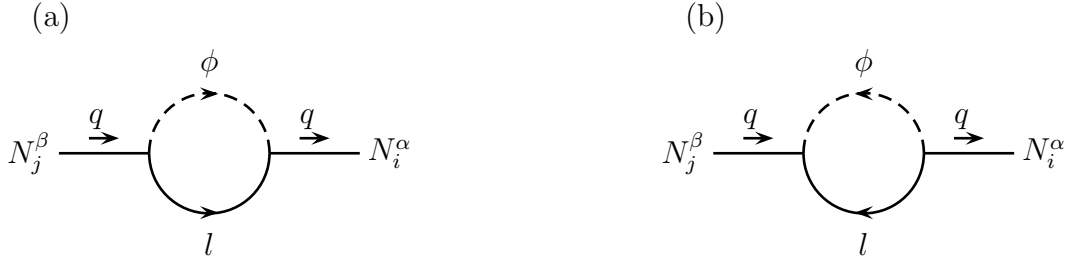}
    \caption{\it Leading order contributions to the self-energy of
      the heavy Majorana neutrinos. \label{fig01}}
  \end{figure}

  At one-loop level the two diagrams in fig.~\ref{fig01} yield the
  self energy 
  \beq
    \S_{\a\b}^{ij}(q)=
    \left(\slash{q}\,P_R\right)_{\a\b}\,\S_R^{ij}(q^2)
       +\left(\slash{q}\,P_L\right)_{\a\b}\,\S_L^{ij}(q^2)\;,
       \label{sigma2}
  \eeq
  where $P_{R,L}={1\over2}(1\pm\g_5)$ are the projectors on right- and
  left-handed chiral states.
  $\S_R$ and $\S_L$ are the contributions of the diagrams
  figs.~(\ref{fig01}a) and (\ref{fig01}b), respectively. They can be
  written as products of a complex function $a(q^2)$ and a hermitian
  matrix $K$,
  \beq
    \S_L(q^2)=\left(\S_R(q^2)\right)^T=a(q^2)\,K\; ,\quad
     K=\l_{\n}^{\dg}\l_{\n}\; .\label{def_aK}
  \eeq
$a(q^2)$ is given by the usual form factor 
$B_0(q^2,0,0)$ \cite{passarino}, whose finite part reads in the
$\Bar{MS}$-scheme, 
  \beq\label{aq2}
a(q^2) = {1\over 16\p^2}\left(\ln{{|q^2|\over \m^2}} - 
          2 - i\p\Theta(q^2)\right)\;.
  \eeq
  For simplicity we will often omit the argument of $a$ in the following,
  however one should keep in mind that $a$ depends on $q^2$.

  According to eqs.~(\ref{S}) and (\ref{sigma2}) the
  resummed propagator $S(q)$ satisfies
  \beq\label{propre}
    \left[\,\slash{q}\,\Big(\,(1-\S_R(q^2))P_R+(1-\S_L(q^2))P_L\,\Big)
    \,-M\,\right]\,S(q)\;=1\;.
    \label{trick1}
  \eeq
  The fermion propagator $S(q)$ consists of four chiral parts
  \beq
    S(q) = P_R\,S^{RR}(q^2)+P_L\,S^{LL}(q^2)
           + P_L\,\slash{q}\,S^{LR}(q^2)
           + P_R\,\slash{q}\,S^{RL}(q^2)\;.
  \eeq  
  Inserting this decomposition into eq.~(\ref{trick1}), and
  multiplying the resulting equation from the left and the right
  with chiral projectors $P_{R,L}$, yields a system of four coupled
  linear equations for the four parts of the propagator. The
  solution reads
  \beqa
    S^{RR}(q^2)&=&\left[(1-\S_L(q^2)){\textstyle q^2\over\textstyle M}
                 (1-\S_R(q^2))-M\right]^{-1}\;,\label{SRR}\\
    S^{LR}(q^2)&=&{1\over M}(1-\S_R(q^2))S^{RR}(q^2)\;,\label{SLR}\\
    S^{LL}(q^2)&=&\left[(1-\S_R(q^2)){\textstyle q^2\over\textstyle M}
                  (1-\S_L(q^2))-M\right]^{-1}\;,\label{SLL}\\
    S^{RL}(q^2)&=&{1\over M}(1-\S_L(q^2))S^{LL}(q^2)\;.\label{SRL}
  \eeqa
  As discussed below, the diagonal elements of $S(q)$ have approximately
  the usual Breit-Wigner form.

  \begin{figure}[t]
    \input{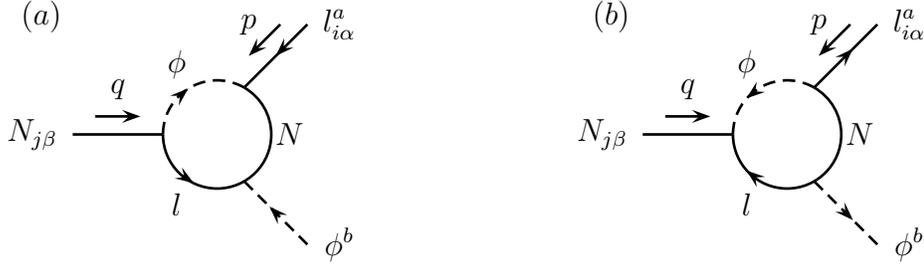}
    \caption{\it One-loop corrections to the couplings of heavy Majorana
  neutrinos $N_j$ to anti-lepton Higgs states (a) and lepton Higgs states (b).
  \label{vert}}
  \end{figure}
\noindent
In addition to the self-energy we need the one-loop vertex function.
The two expressions for the coupling of $N$ to $\bar{l},\bar{\f}$ (fig.~2a) 
and $N$ to $l,\f$ (fig.~2b) can be written as
\beqa
\bar{\G}^{ji}_{\b \a,a b}(q,p)&=&+i\e_{ab}\left((K M b(q,p) \l_\n^T)_{ji} q_\m
 + (K M c(q,p) \l_\n^T)_{ji} p_\m\right) (C \g^\m P_L)_{\b\a}\;,\label{gam}\\
\G^{ij}_{\a \b,a b}(q,p)&=&-i\e_{ab}\left((\l_\n^* M b(q,p) K)_{ij} q_\m
 + (\l_\n^* M c(q,p) K)_{ij} p_\m\right) (P_R \g^\m)_{\a\b}\; .\label{gamb}
\eeqa
Here $b(q,p)$ and $c(q,p)$ are diagonal matrices whose elements are given
by the standard form factors $C_0$ and $C_{12}$ \cite{passarino},
\beqa
b_k(q,p)&=&{1\over 16\p^2}\left(C_0(-p-q,q,M_k,0,0)
           +C_{12}(-p-q,q,M_k,0,0)\right)\;,\label{bk}\\
c_k(q,p)&=&{1\over 16\p^2}\left(C_0(-p-q,q,M_k,0,0)
           + 2 C_{12}(-p-q,q,M_k,0,0)\right)\;.
\eeqa
Since we shall only consider amplitudes with massless on-shell leptons,
the terms proportional to $c_k$ will not contribute. We shall only need
the imaginary part of $b_k$ which is given by
\beq
\mbox{Im}\{b_k(q^2)\} = {1\over 16\p \sqrt{q^2}M_k} 
     f\left({M_k^2\over q^2}\right) \Theta(q^2)\;,\label{imbk}
\eeq
where the function $f$ is defined as
\beq\label{fx}
f(x) = \sqrt{x}\left(1-(1+x)\ln\left({1+x\over x}\right)\right)\;.
\eeq

\noindent
{\bf\large Transition matrix elements}\\
  
The two lepton-number violating and the two lepton-number conserving 
processes are shown in figs.~(3a)~-~(3d). Consider first the contributions
of the full propagator, where the full vertices are replaced by tree
couplings. The four scattering amplitudes read
\beqa\label{2bamp}
\langle \bar{l}^d_j(p') \bar{\f}^e(q-p')|l^a_i(p) \f^b(q-p)\rangle
&=& +i \e_{ab}\e_{de} (\l^T_\n)_{lj}(\l^T_\n)_{ki}\NO\\
&&\qquad (C P_L v(p'))^T\ S^{LL}_{lk}(q)C^{-1}\ (C P_L u(p))\; ,\\
\langle l^d_j(p') \f^e(q-p')|\bar{l}^a_i(p) \bar{\f}^b(q-p)\rangle
&=& +i \e_{ab}\e_{de} (\l^{\dg}_\n)_{lj}(\l^{\dg}_\n)_{ki}\NO\\
&&\qquad (\bar{u}(p') P_R)\ S^{RR}_{lk}(q)C^{-1}\ (\bar{v}(p)P_R)^T\; ,\\
\langle l^d_j(p') \f^e(q-p')|l^a_i(p) \f^b(q-p)\rangle
&=& -i \e_{ab}\e_{de} (\l^{\dg}_\n)_{lj}(\l^T_\n)_{ki}\NO\\
&&\qquad (\bar{u}(p') P_R)\ S^{RL}_{lk}(q)\slash{q}C^{-1}\ (C P_L u(p))\; ,\\
\langle \bar{l}^d_j(p') \bar{\f}^e(q-p')|\bar{l}^a_i(p) \bar{\f}^b(q-p)\rangle
&=& -i \e_{ab}\e_{de} (\l^T_\n)_{lj}(\l^{\dg}_\n)_{ki}\NO\\
&&\qquad (C P_L v(p'))^T \ S^{LR}_{lk}(q)\slash{q}C^{-1}\ 
(\bar{v}(p) P_R)^T\, .
\eeqa
Here $a,b,d,e$ denote the SU(2) indices of lepton and Higgs fields and
$i,j,k,l$ are generation indices. The relative signs follow from Fermi 
statistics.
 
We are particularly interested in the contributions of a single heavy
neutrino to the scattering amplitudes. In order to determine these
contributions we have to find the poles and the residues of the
propagator matrix. Here an unfamiliar complication arises due to the fact
that the self-energy matrix is different for left- and right-handed states.
Hence, the different chiral projections of the propagator matrix are
diagonalized by different matrices. 

$S^{LL}$ and $S^{RR}$ are symmetric complex matrices, since 
$\S_L(q^2)=\left(\S_R(q^2)\right)^T$. Hence, $S^{LL}$ and $S^{RR}$ can
be diagonalized by complex orthogonal matrices $V$ and $U$, respectively,
\beq\label{sdiag}
S^{LL}(q^2) = V^T(q^2) M D(q^2) V(q^2)\;,\quad
S^{RR}(q^2) = U^T(q^2) M D(q^2) U(q^2)\;.
\eeq
Splitting the self-energy into a diagonal and an off-diagonal part, 
\beq
\S_L(q^2) = \S_D(q^2) + \S_N(q^2)\;,
\eeq
one finds
\beq\label{dresum}
D^{-1}(q^2) = q^2(1-\S_D(q^2))^2 - M^2 + \co(\S_N^2)\;.
\eeq
One can easily identify real and imaginary parts of the propagator poles.
The pole masses are given by
\beq\label{polem}
\Bar{M}_i^2 = Z_{Mi} M_i^2\; ,\quad Z_{Mi} = \left(1 + {K_{ii}\over 8\p^2}
\left(\ln{M_i^2\over \m^2} - 2\right)\right)\;,
\eeq
and the widths are $\G_i = K_{ii}M_i/(8\p)$. In the vicinity of the poles
the propagator has the familiar Breit-Wigner form
\beq
D_i(q^2) \simeq {Z_i\over q^2 - \Bar{M}_i^2 + i \Bar{M}_i \G_i}\;, \quad
Z_i = \left(1 + {K_{ii}\over 8\p^2}\left(\ln{M_i^2\over\m^2}-1\right)\right)\;.
\eeq 
 
 \begin{figure}[t]
    \input{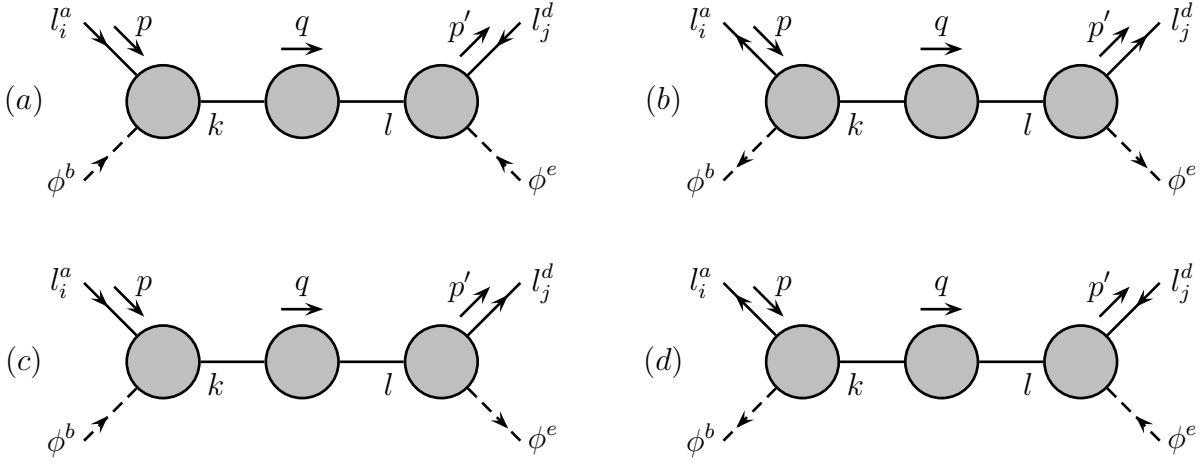}
    \caption{\it s-channel contributions to lepton-Higgs scattering, 
     including full propagators and vertices. \label{schannel}}
 \end{figure}

We can now easily write down the contribution of a single resonance $N_l$ 
with spin $s$ to the  lepton-Higgs scattering amplitudes.  
Suppressing spin indices for massless fermions, one has
\beqa
\langle \bar{l}^d_j(p') \bar{\f}^e(q-p')|l^a_i(p) \f^b(q-p)\rangle_l
&=& \langle \bar{l}^d_j(p') \bar{\f}^e(q-p')|N_l(q,s)\rangle \NO\\
&&\quad i D_l(q^2)\ \langle N_l(q,s)|l^a_i(p) \f^b(q-p)\rangle\;,\label{lbl}\\
\langle l^d_j(p') \f^e(q-p')|\bar{l}^a_i(p) \bar{\f}^b(q-p)\rangle_l
&=&\langle l^d_j(p') \f^e(q-p')|N_l(q,s)\rangle \NO\\
&&\quad i D_l(q^2)\ \langle N_l(q,s)|\bar{l}^a_i(p)\bar{\f}^b(q-p)\rangle\;,\\
\langle l^d_j(p') \f^e(q-p')|l^a_i(p) \f^b(q-p)\rangle_l
&=&\langle l^d_j(p') \f^e(q-p')|N_l(q,s)\rangle_{LC} \NO\\
&&\quad i D_l(q^2)\ \langle N_l(q,s)|l^a_i(p) \f^b(q-p)\rangle\;,\\
\langle\bar{l}^d_j(p')\bar{\f}^e(q-p')|\bar{l}^a_i(p)\bar{\f}^b(q-p)
                 \rangle_l
&=&\langle\bar{l}^d_j(p')\bar{\f}^e(q-p')|N_l(q,s)\rangle_{LC} \NO\\
&&\quad i D_l(q^2)\ \langle N_l(q,s)|\bar{l}^a_i(p)\bar{\f}^b(q-p)\rangle\;.
\eeqa
Here the subscript $LC$ distinguishes an amplitude defined by a 
lepton-number conserving process from the same amplitude defined by a
lepton-number violating process. From eqs.~(\ref{SRR})~-~(\ref{SRL}) and 
(\ref{sdiag}) one finds
\beqa
\langle N_l(q,s)| l^a_i(p) \f^b(q-p)\rangle
&=& +i \e_{ab}\left(V(q^2)\l_\n^T\right)_{li}\ \bar{u}_s(q,M_l)P_L u(p)\;,
\label{ln}\\
\langle \bar{l}^d_j(p') \bar{\f}^e(q-p')|N_l(q,s)\rangle 
&=&-i\e_{de}\left(\l_\n V^T(q^2)\right)_{jl}\bar{v}_s(q,M_l)P_L v(p')\;, 
\label{nlc}\\
\langle N_l(q,s)|\bar{l}^a_i(p) \bar{\f}^b(q-p)\rangle 
&=&-i \e_{ab}\left(U(q^2)\l_\n^{\dg}\right)_{li}\ \bar{v}(p)P_R v_s(q,M_l)\;,
\label{lcn}\\
\langle l^d_j(p') \f^e(q-p')|N_l(q,s)\rangle 
&=&+i\e_{de}\left(\l_\n^* U^T(q^2)\right)_{jl}\bar{u}(p')P_R u_s(q,M_l)\;,
\label{nl}\\
\langle l^d_j(p') \f^e(q-p')|N_l(q,s)\rangle_{LC} 
&=&+i\e_{de}\left(\l_\n^* {1\over M} (1 - \S_L(q^2)) V^T(q^2) M\right)_{jl}
\NO\\ &&\hspace{1.2cm}\bar{u}(p')P_R u_s(q,M_l)\;,
\label{nlLC}\\
\langle\bar{l}^d_j(p')\bar{\f}^e(q-p')|N_l(q,s)\rangle_{LC}
&=&-i\e_{de}\left(\l_\n {1\over M} (1 - \S_R(q^2)) U^T(q^2) M\right)_{jl}
\NO\\&&\hspace{1.2cm}\bar{v}_s(q,M_l)P_L v(p')\;, \label{nlcLC}
\eeqa
where we have used the identity $C \bar{v}^T_s(p) = u_s(p)$.

Eqs.~(\ref{ln}) and (\ref{nlc}) describe the coupling of the Majorana field
$N$ to the lepton fields $l_i$ and the Higgs field $\f$, and eqs.~(\ref{lcn})
and (\ref{nl}) give the couplings of $N$ to the charge conjugated fields
$\bar{l}_i$ and $\f^*$. In the case of CP conservation, one has $\l_{\n ij}=
\l_{\n ij}^*$, which implies $K=K^T$ and therefore
\beq
\S_L(q^2) = \S_R(q^2)\;, \quad V(q^2) = U(q^2)\; .
\eeq
This yields
\beq
\langle N_l(q,s)| l^a_i(p) \f^b(q-p)\rangle 
= \langle N_l(\tilde{q},s)|\bar{l}^a_i(\tilde{p}) 
                           \bar{\f}^b(\tilde{q}-\tilde{p})\rangle\; ,
\eeq
with $\tilde{q}=(q_0,-\vec{q})$, $\tilde{p}=(p_0,-\vec{p})$,
as required by CP invariance.

The amplitudes given in eqs.~(\ref{ln}) - (\ref{nl}) have been obtained
from the lepton-number violating processes figs.~(3a) and (3b). The 
lepton-number conserving processes figs.~(3c) and (3d) yield the amplitudes
given in eqs.~(\ref{nlLC}) and (\ref{nlcLC}). The consistent definition of
an on-shell contribution of a single heavy Majorana neutrino to  the
two-body scattering amplitudes requires that the transition amplitudes 
extracted from lepton-number conserving and lepton-number violating
processes are consistent. This implies
\beqa  
\langle l^d_j(p') \f^e(q-p')|N_l(q,s)\rangle
&=&\langle l^d_j(p') \f^e(q-p')|N_l(q,s)\rangle_{LC}\; ,\\ 
\langle\bar{l}^d_j(p')\bar{\f}^e(q-p')|N_l(q,s)\rangle
&=&\langle\bar{l}^d_j(p')\bar{\f}^e(q-p')|N_l(q,s)\rangle_{LC}\; .
\eeqa
From eqs.~(\ref{nlc}), (\ref{nl}), (\ref{nlLC}) and (\ref{nlcLC}) it
is clear that these relations are fulfilled if the mixing matrices
$V(q^2)$ and $U(q^2)$ satisfy certain consistency relations. Assuming that
the matrix $\l_\n$ has an inverse, one reads off
\beqa
U_{ij}(M_i^2) &=& \left(M V(M_i^2) (1-\S_R(M_i^2)){1\over M}\right)_{ij}\; ,
\label{consU}\\
V_{ij}(M_i^2) &=& \left(M U(M_i^2) (1-\S_L(M_i^2)){1\over M}\right)_{ij}\; .
\label{consV}
\eeqa

The matrices $V$ and $U$ are determined by the requirement that the
expressions (cf.~eqs.~(\ref{sdiag}))
\beqa
V(q^2)(S^{LL}(q^2))^{-1}V^T(q^2) = V(q^2)\left(
(1-\S_R(q^2)){q^2\over M}(1-\S_L(q^2)) - M\right)V^T(q^2)\;,\\
U(q^2)(S^{RR}(q^2))^{-1}U^T(q^2) = U(q^2)\left(
(1-\S_L(q^2)){q^2\over M}(1-\S_R(q^2)) - M\right)U^T(q^2)\;,
\eeqa
are diagonal on-shell, i.e., at $q^2=M_i^2$. Using $\S_L = \S_D + \S_N$, 
and writing
\beqa
V(q^2) &=& 1 + v(q^2)\; ,\quad v(q^2)=-v^T(q^2)\; ,\label{Vv}\\
U(q^2) &=& 1 + u(q^2)\; ,\quad u(q^2)=-u^T(q^2)\; ,\label{Uu}
\eeqa
a straightforward calculation yields
\beqa
v_{ij}(q^2)&=&w_{ij}(q^2)\left(M_i \S_{Nji}(q^2) + M_j \S_{Nij}(q^2)\right)\;,
\label{vw}\\
u_{ij}(q^2)&=&w_{ij}(q^2)\left(M_i \S_{Nij}(q^2) + M_j \S_{Nji}(q^2)\right)\;,
\label{uw}
\eeqa
where
\beq\label{ww}
w_{ij}(q^2)^{-1} = (M_i-M_j)\left(1 + {M_i M_j\over q^2}\right)
 - 2 a(q^2) \left(M_i K_{jj} - M_j K_{ii}\right)\;.
\eeq
These equations give the matrices $V$ and $U$ to leading order in $\S_N$.
They are meaningful as long as the matrix elements of $\S_N$ are small
compared to those of $w^{-1}$. 

Inserting eqs.~(\ref{vw}) and (\ref{uw}) in eqs.~(\ref{consU}) and 
(\ref{consV}), one finds that 
the consistency conditions for the mixing matrices $V$ and $U$ are fulfilled
to leading order in $\S_N$. We conclude that the contribution of a single
heavy neutrino to  two-body scattering processes can indeed be
consistently defined. The pole masses are given by eq.~(\ref{polem}) and the
couplings to lepton-Higgs initial and final states are given by 
eqs.~(\ref{ln}) - (\ref{nl}).\\
 
\noindent
{\bf\large CP asymmetry in heavy neutrino decays}\\

It is now straightforward to evaluate the CP asymmetry in the decay of a
heavy Majorana neutrino,
\beq\label{cpasym}
\e_i = {\G(N_i\rightarrow l\f) - \G(N_i\rightarrow \bar{l}\bar{\f})\over 
        \G(N_i\rightarrow l\f) + \G(N_i\rightarrow \bar{l}\bar{\f})} \; .
\eeq 
From eqs.~(\ref{nlc}) and (\ref{nl}) one obtains for the partial decay widths,
including mixing effects,
\beqa
\G_M(N_i\rightarrow \bar{l}\bar{\f}) &\propto& 
\sum_j |(\l_\n V^T(M_i^2))_{ji}|^2 \;,\label{gnbarl}\\
\G_M(N_i\rightarrow l\f) &\propto& \sum_j |(\l_\n^* U^T(M_i^2))_{ji}|^2 \;.
\label{gnl}
\eeqa
To leading order in $\l_\n^2$ this yields the asymmetry 
(cf.~eqs.~(\ref{Vv}), (\ref{Uu})),
\beq
\e_i^M = {1\over K_{ii}} \mbox{Re}\left\{(u(M_i^2)K)_{ii} -
                                (v(M_i^2)K^T)_{ii}\right\}\;.
\eeq
Using eqs.~(\ref{vw}) - (\ref{ww}) and (\ref{aq2}), one finally obtains
\beq\label{cpan}
\e_i^M = -{1\over 8\p}\sum_j |w_{ij}(M_i^2)|^2 (M_i^2-M_j^2)
{M_j\over M_i}  {\mbox{Im}\{K_{Nij}^2\}\over K_{ii}}\; .
\eeq

Consider first the case where differences between heavy neutrino masses
are large, i.e., $|M_i - M_j|\gg |\G_i - \G_j|$. Eq.~(\ref{cpan}) then
simplifies to
\beq\label{cpwave}
\e_i^M = - {1\over 8\p}\sum_j {M_i M_j \over M_i^2 - M_j^2}
        {\mbox{Im}\{K_{Nij}^2\}\over K_{ii}}\; .
\eeq
This is the familiar CP asymmetry due to flavour mixing \cite{flanz}. It has
previously been obtained by considering directly the self-energy correction
to the Majorana neutrino decay, without any resummation. The CP asymmetry
$\e_i$ reaches its maximum for $|M_i - M_j|\sim |\G_i - \G_j|$, where
the perturbative expansion breaks down. 

Interesting is also the limiting case where the heavy neutrinos become
mass degenerate. From eq.~(\ref{cpan}) it is obvious that the CP asymmetry
vanishes in this limit. The vanishing of the CP asymmetry for mass 
degenerate heavy
neutrinos is expected on general grounds, since in this case the CP
violating phases of the matrix $K$ can be eliminated by a change of basis.
 
The CP asymmetry due to the vertex corrections is easily obtained using 
eqs.~(\ref{gam}), (\ref{gamb}), (\ref{imbk}) and (\ref{fx}).
The partial decay widths corresponding to the full vertex read
\beqa
\G_V(N_i\rightarrow \bar{l}\bar{\f}) &\propto& 
\sum_j |(\l_\n (1 - M b K^T M))_{ji}|^2 \;,\label{gvbarl}\\
\G_V(N_i\rightarrow l\f) &\propto& \sum_j |(\l_\n^* (1 - M b K M))_{ji}|^2 \;.
\label{gvl}
\eeqa
For the corresponding CP asymmetry (\ref{cpasym}) one obtains the familiar 
result
\beq
\e_i^V =-{1\over 8\p}\sum_j {\mbox{Im}\{K_{Nij}^2\}\over K_{ii}} 
         f\left({M_j^2\over M_i^2}\right) \; ,
\eeq
where the function $f(x)$ has been defined in eq.~(\ref{fx}).\\

\noindent   
{\bf\large CP asymmetries in two-body processes}\\

Let us now consider the CP asymmetries in two-body processes. Here we have
to take into account the s-channel amplitudes shown in figs.~(3a) and (3b),
with vertex functions up to one-loop, and the two u-channel amplitudes
depicted in figs.~(4a) and (4b). For the u-channel amplitudes vertex and
self-energy corrections can be omitted to leading order
since the absorptive parts vanish.
 
 \begin{figure}[t]
    \input{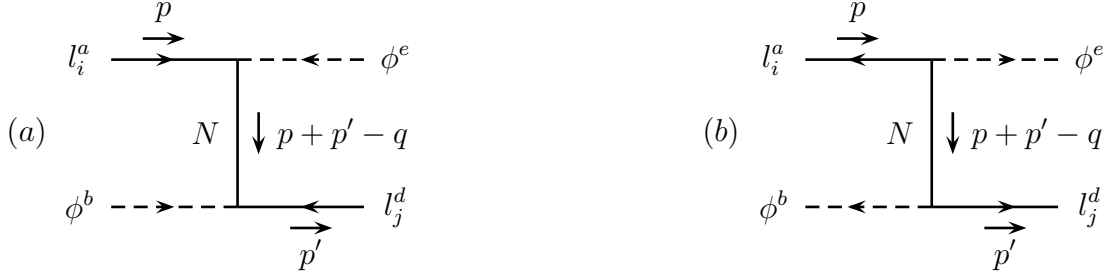}
    \caption{\it u-channel contributions to lepton-Higgs scattering. 
     \label{uchannel}}
 \end{figure}
 
In the following we shall evaluate various contributions to the CP asymmetry
\beq
\e \equiv {\Delta |\cm|^2\over 2 |\cm |^2}
   \equiv {|\cm (\bar{l}\bar{\f} \rightarrow l\f)|^2 -
      |\cm (l\f \rightarrow \bar{l}\bar{\f})|^2 \over
      |\cm (\bar{l}\bar{\f} \rightarrow l\f)|^2 +
      |\cm (l\f \rightarrow \bar{l}\bar{\f})|^2} \;,
\eeq
where we always sum over generations in initial and final states. There
are contributions from the full s-channel propagator, $\Delta |\cm|^2_s$,
from the interference between s-channel amplitudes at tree-level and with
one-loop vertex corrections, $\Delta |\cm|^2_{s,\G}$, the interference between 
tree-level s-channel and u-channel amplitudes, $\Delta |\cm|^2_{s,u}$,
and the interference between s-channel with one-loop vertex corrections and 
u-channel amplitudes, $\Delta |\cm|^2_{u,\G}$.

Consider first the CP asymmetry $\e_s$ due to the full propagator. The 
contribution of a single intermediate neutrino $N_i$ is (cf.~(\ref{lbl}),
(\ref{ln}),(\ref{nlc})) 
\beq
|\cm_i(l\f \rightarrow \bar{l}\bar{\f})|^2_s \propto |D_i(q^2)|^2
\sum_j|(V(q^2)\l_\n^T)_{ij}|^2\  \sum_k|(\l_\n V^T(q^2))_{ki}|^2\; .
\eeq
Comparison with eq.~(\ref{gnbarl}) yields immediately
\beq
|\cm_i(l\f \rightarrow \bar{l}\bar{\f})|^2_s \propto |D_i(q^2)|^2
\G_M(N_i \rightarrow \bar{l}\bar{\f})^2\; .
\eeq
Similarly, one has for the charge conjugated process 
\beq
|\cm_i(\bar{l}\bar{\f} \rightarrow l\f)|^2_s \propto |D_i(q^2)|^2
\G_M(N_i\rightarrow l\f)^2\; .
\eeq
The corresponding CP asymmetry is, as expected, twice the asymmetry
in the decay due to mixing,
\beq
\e_s^{(i)} = {\Delta |\cm_i|^2_s \over 2 |\cm_i|^2_s}\simeq 2 \e^M_i\; .
\eeq

It is very instructive to compare the contribution of a single resonance
with the CP asymmetry $\e_s$ for the full propagator. Due to the structure 
of the propagators $S^{LL}$ and $S^{RR}$ it is difficult to evaluate $\e_s$ 
exactly. However, one may easily calculate $\e_s$ perturbatively in powers 
of $\S_N$, like 
the mixing matrices $V(q^2)$ and $U(q^2)$ in the previous section.

The full propagator (cf.~(\ref{propre})) reads to first order in $\S_N$,
\beq
S(q)=S_D(q) + S_D(q) \slash{q}[\S_N^T(q^2)P_R + \S_N(q^2)P_L] S_D(q)+\ldots\;,
\eeq
where (cf.~(\ref{dresum}))
\beq
S_D(q) = [\slash{q}(1-\S_D(q^2)) + M] D(q^2)\;. 
\eeq

It is now straightforward to calculate the matrix elements of two-body 
processes, summed over generations in initial and final states,
\beqa
|\cm (l\f \rightarrow \bar{l}\bar{\f})|^2_s &=&
16\ p\cdot p'\ q^2 \left({1\over 2 q^2} 
\mbox{tr}[K M D(q^2) K^T M D^*(q^2)]+\right.\NO\\
&& \mbox{Re} \left\{\mbox{tr} [K M D(q^2) \S_N^T(q^2)
(1-\S_D(q^2)) D(q^2) K^T M D^*(q^2) + \right. \NO\\
&&\left. K D(q^2) (1-\S_D(q^2))\S_N(q^2)
M D(q^2) K^T M D^*(q^2)]\right\} + \ldots\Bigg)\, .
\eeqa
This yields for the sum and the difference of the CP conjugated matrix 
elements,
\beqa
2 |\cm|^2_s &=& 16\ p\cdot p' \sum_{i,j} A_{ij}+\ldots\;,\\
\Delta |\cm|^2_s &=&- 16\ p\cdot p' \sum_{i,j}(B_{ij}+C_{ij})+\ldots\;,
\eeqa
where
\beqa
A_{ij}&=&\mbox{Re}\left\{K_{ij}^2 M_i M_j D_j(q^2) D_i^*(q^2)\right\}\;,
\label{aij}\\
B_{ij}&=& i \mbox{Im}\{K_{Nij}\}^2 M_i M_j D_j(q^2) D_i^*(q^2)\;,\label{bij}\\
C_{ij}&=& 4 q^2 \mbox{Re}\left\{ia(q^2) \mbox{Im}\{K_{Nij}^2\} M_i M_j
(1-\S_D(q^2)_i)K_{ii}D_j(q^2) |D_i(q^2)|^2\right\}\;.\label{cij}
\eeqa

For $q^2 \simeq M_i^2$ the expressions $A_{ij}$ and $C_{ij}$ are dominated
by the contribution of a single resonance $N_i$,
\beqa
A_{ii} &\simeq& K_{ii}^2 M_i^2 |D_i(q^2)|^2 \;,\label{Apole}\\
C_{ij} &\simeq& {1\over 4\pi} \mbox{Im}\{K_{Nij}^2\} {M_i^3 M_j\over 
                 M_i^2-M_j^2} K_{ii} |D_i(q^2)|^2\;. \label{Cpole}
\eeqa
From eqs.~(\ref{cpwave}), (\ref{Apole}) and (\ref{Cpole}) one reads off
that the sum over $C_{ij}$ yields precisely the contribution of the
resonance $N_i$ to the CP asymmetry,
\beq
-{\sum_j C_{ij}\over A_{ii}} = 
-{1\over 4\p}\sum_j {M_i M_j \over M_i^2 - M_j^2}
        {\mbox{Im}\{K_{Nij}^2\}\over K_{ii}} =2 \e_i^M\;.
\eeq

The second contribution to the CP asymmetry $\e_s$ is due to the sum over 
$B_{ij}$ (cf.~eq.~(\ref{bij})). $B_{ij}$ involves two different propagators 
($i\neq j$) and corresponds to an interference term. Using $D_j^{*-1}(q^2)=
q^2-M_j^2-2a^*(q^2) q^2 K_{jj}$ and $2 q^2 \mbox{Im}\{a(q^2)\} K_{ii}=
-\mbox{Im}\{D_i^{-1}(q^2)\}$, one can rewrite $C_{ij}$ to leading order in
$\l_\n^2$ as follows,
\beq\label{cijre}
C_{ij} = -i \mbox{Im}\{K_{Nij}^2\} M_i M_j D_j^{*-1}(q^2) D_i^{-1}(q^2) 
        |D_i(q^2)|^2 |D_j(q^2)|^2\;.
\eeq
Comparing eqs.~(\ref{bij}) and (\ref{cijre}) it is obvious that the sum of both
terms, i.e., the CP asymmetry $\e_s$ corresponding to the full propagator, is 
identically zero! The pole contribution is cancelled by the interference of 
the pole term with an off-shell propagator.

The contribution to the CP asymmetry $\Delta |\cm|^2_{s,\G}$ can be
computed in a similar manner. The diagrams fig.~(3a) and (3b) yield two
contributions for the two vertices. After some algebra one obtains the
result (cf.~(\ref{bk}))
\beqa
\Delta |\cm|^2_{s,\G} &=& - 64\ p\cdot p'\ q^2 \sum_{i,j,k} D_{ijk} 
+ \ldots\;,\label{dmgs} \\
D_{ijk} &=& \mbox{Im}\{K_{ik}K_{jk}K_{ij}\}
\mbox{Im}\{b_k(q^2)\} M_k M_j
D_i(q^2) D_j^*(q^2)\;.\label{dijk}
\eeqa
For $q^2\simeq M_i^2$, one reads off that the sum over $D_{ijk}$ yields,
as expected, twice the vertex CP asymmetry,
\beq
\e_{s,\G}(M_i^2) \simeq  {\sum_k D_{iik}\over A_{ii}} = 
-{1\over 4\p}\sum_k {\mbox{Im}\{K_{Nik}^2\}\over K_{ii}} 
         f\left({M_k^2\over M_i^2}\right) = 2 \e_i^V\; .
\eeq
A result very similar to eqs.~(\ref{dmgs}), (\ref{dijk}) is obtained for
the asymmetry $\Delta|\cm |^2_{s,u}$, the interference between tree-level
s-channel and u-channel amplitudes. One finds ($u=(q-p-p')^2$), 
\beqa
\Delta |\cm|^2_{s,u} &=& - 32\ p\cdot p'\ q^2 \sum_{i,j,k} E_{ijk} + 
\ldots\;,\label{emgs} \\
E_{ijk} &=& \mbox{Im}\{K_{ik}K_{jk}K_{ij}\}
\mbox{Im}\{a(q^2)\} M_k M_j
D_i(q^2) D_j^*(q^2) D_k^*(u)\;.\label{eijk}
\eeqa
Integrating the expressions over phase space and using
\beq
\int_{-q^2}^0 du {2 p\cdot p'\over u - M_k^2}
= {q^2 \sqrt{q^2}\over M_k} f\left({M_k^2\over q^2}\right)\; ,
\eeq
one finds the cancellation
\beq
\int_{-q^2}^0 du (\Delta |\cm|^2_{s,\G} + \Delta |\cm|^2_{s,u}) = 0\; .
\eeq

Finally, we have to consider the CP asymmetry $\Delta |\cm|^2_{u,\G}$. A
straightforward calculation yields
\beqa
\Delta |\cm|^2_{u,\G} &=& - 32\ p\cdot p'\ q^2 \sum_{i,j,k} F_{ijk} 
+ \ldots\;,\label{fmgs} \\
F_{ijk} &=& \mbox{Im}\{K_{ik}K_{jk}K_{ji}\}
\mbox{Im}\{b_k(q^2)\} M_k M_j D_i(q^2) D_j^*(u) \;.\label{fijk}
\eeqa
After integration over $u$ the resulting matrix $\bar{F}_{ijk}$ is
antisymmetric in the indices $j$ and $k$. As a consequence, the asymmetry
$\Delta |\cm|^2_{u,\G}$ is identically zero.

As we have seen, the total CP asymmetry vanishes to leading order in 
$\l_\n^2$. This result has previously been obtained in \cite{rcv}. It
follows from unitarity and CPT invariance. The considered T-matrix elements
satisfy the unitarity relation
\beq\label{imt}
2\ \mbox{Im}\langle l\f|T|l\f\rangle = \langle l\f|T^\dagger T|l\f\rangle\;.
\eeq
If, in perturbation theory, the leading contribution to the right-hand side
is given by two-particle intermediate states, one has
\beq\label{twpst}
\sum_l \langle l\f|T^\dagger T|l\f\rangle = \sum_{l,l'}\left(
|\langle l'\f|T|l\f\rangle|^2 +|\langle \bar{l'}\bar{\f}|T|l\f\rangle|^2 
\right) + \ldots\;.
\eeq
CPT invariance implies
\beq\label{cpt}
\langle l'\f|T|l\f\rangle = 
\langle \bar{l}\bar{\f}|T|\bar{l'}\bar{\f}\rangle\;.
\eeq
From eqs.~(\ref{imt})~-~(\ref{cpt}) one then obtains
\beq
\sum_{l,l'}\left(
|\langle \bar{l'}\bar{\f}|T|l\f\rangle|^2 -
|\langle l'\f|T|\bar{l}\bar{\f}\rangle|^2 \right) + \ldots = 0 \;.\label{cpa}
\eeq
In \cite{rcv} it was concluded that away from resonance poles, where ordinary
perturbation theory holds, the CP asymmetry (\ref{cpa}) vanishes to order
$\l_\n^6$. Corrections due to four-particle intermediate states are
$\co (\l_\n^8)$. In this paper we have developed a resummed perturbative 
expansion in powers of $\S_N$ which is also valid for $s\simeq M_i^2$. The 
same argument then implies that in this case the CP asymmetry (\ref{cpa}) 
vanishes to order $\l_\n^2$ with corrections $\co (\l_\n^4)$.

The nature of the cancelation is different for different subprocesses. For
the full propagator, the CP asymmetry vanishes identically for fixed
external momenta. Interference contributions between various s-channel
and u-channel amplitudes cancel after phase space integration. In
applications at finite temperature the standard practice \cite{wolfram} is
to treat in the Boltzmann equations resonance contributions and the remaining 
two-body cross sections differently. This procedure yields for the CP 
asymmetry of the decaying heavy neutrino $N_i$ the sum of mixing and vertex 
contribution, $\e_i = \e_i^M + \e_i^V$. However, the generation of a
lepton asymmetry is an out-of-equilibrium process and one may worry to what
extent the result is affected by interference terms which are neglected.
It therefore appears important to study systematically corrections to
the Boltzmann equations \cite{yosh}.\\

We would like to thank E.~Roulet and F.~Vissani for clarifying discussions
and comments. 


\end{document}